\documentclass{PoS}
\usepackage{amsmath}
\usepackage{slashed}
\usepackage{subcaption}
\usepackage{enumitem}

\usepackage[dvipsnames]{xcolor}
\definecolor{hybrid}{rgb}{0.5, 0.3, 0}
\definecolor{composite}{rgb}{0, 0, 1}
\definecolor{vector}{rgb}{0.5, 0, 0.5}
\definecolor{pseudoscalar}{rgb}{0, 0.8, 0}
\definecolor{scalar}{rgb}{1, 0, 0}

\title{Lattice Analysis of $SU(2)$ with 1 Adjoint Dirac Flavor}

\ShortTitle{$SU(2)$ Adjoint}

\author{Zhen Bi \\
        Massachusetts Institute of Technology \\
        E-mail: \email{zbi@mit.edu}}

\author{\speaker{Anthony Grebe}\\
        Massachusetts Institute of Technology\\
        E-mail: \email{agrebe@mit.edu}}
        
\author{Gurtej Kanwar \\
        Massachusetts Institute of Technology \\
        E-mail: \email{gurtej@mit.edu}}
        
\author{Patrick Ledwith \\
        Massachusetts Institute of Technology \\
        Harvard University \\
        E-mail: \email{pled@mit.edu}}
        
\author{David Murphy \\
        Massachusetts Institute of Technology \\
        E-mail: \email{djmurphy@mit.edu}}
        
\author{Michael L. Wagman \\
        Massachusetts Institute of Technology \\
        E-mail: \email{mlwagman@mit.edu}}

\abstract{Recently $SU(2)$ Yang-Mills theory with one massless adjoint Dirac quark flavor emerges as a novel critical theory that can describe the evolution between a trivial insulator and a topological insulator in AIII class in $3+1$ dimensions. There are several classes of conjectured infrared dynamics for this theory. One possibility is that the theory undergoes spontaneous chiral symmetry breaking, with two massless Goldstone bosons (the scalar diquark and its antiparticle) in the infrared. Another scenario, which is suggested by previous lattice studies by Athenodorou et al., is that the IR sector of the theory is a strongly interacting conformal field theory as the quark mass vanishes.  The most recent theoretical proposals argue for a case that in the infrared a composite fermion composed of two quarks and an antiquark becomes massless and non-interacting as the quark mass goes to zero, while other sectors are decoupled from this low-energy fermion. This work expands upon previous studies by including the composite fermion to investigate which of these three potential scenarios captures the infrared behavior of this theory.}

\FullConference{37th International Symposium on Lattice Field Theory - Lattice2019\\
		16-22 June 2019\\
		Wuhan, China}

\begin{document}

\section{Motivation}
Ref. \cite{senthil} recently proposed that the phase transition between a trivial insulator and a topological insulator of the AIII class in $3+1$ dimensions, while describable by a theory of a single massless non-interacting Dirac fermion, can also be described by an emergent $3 + 1$ dimensional $SU(2)$ gauge theory with one adjoint massless Dirac fermion (referred to below as massless $SU(2)$ adjoint). The infrared physics of $SU(2)$ adjoint has also been recently studied in Refs.~\cite{athenodorou-1,athenodorou-2,Anber:2018tcj,scenarios,Wan:2018djl} from a variety of perspectives, and several scenarios appear as possible descriptions of the low-energy physics of $SU(2)$ adjoint as summarized in Ref~\cite{Wan:2018djl}.

Constraints from anomaly matching demand that any low-energy effective field theory (EFT) of massless $SU(2)$ adjoint include massless degrees of freedom~\cite{senthil,Anber:2018tcj,scenarios,Wan:2018djl}. One possibility, motivated by studying relevant deformations of $\mathcal{N}=2$ supersymmetric Yang-Mills theory~\cite{scenarios}, is that the low-energy EFT of massless $SU(2)$ adjoint includes spontaneous chiral symmetry breaking with massless Goldstone bosons or an emergent gauge field. Another scenario is that the $SU(2)$ adjoint is inside the conformal window \cite{adjoint-1/2,adjoint-3/2} and is described by a strongly interacting conformal field theory in the infrared~\cite{athenodorou-1}. Finally, very recent proposals advanced in several theoretical studies \cite{senthil,Anber:2018tcj,scenarios,Wan:2018djl} indicate a novel situation where the low energy theory consists of a free massless Dirac fermion augmented by another decoupled sector.\footnote{A fully (1-form and 0-form) symmetric topological theory for the decoupled sector is excluded by a recent discussion in Ref.~\cite{Cordova:2019bsd}. Nonetheless, scenarios with broken symmetries for the decoupled sector are still possible \cite{scenarios,Wan:2018djl}. The decoupled sector can be either gapped (a $Z_2$ TQFT that spontaneously breaks the axial symmetry) or gapless (a $U(1)$ photon theory that spontaneously breaks the 1-form symmetry) \cite{scenarios}.}

Previous lattice studies, predating the Dirac fermion conjecture, found evidence that suggests the IR dynamics of the theory is likely lying inside or near the onset of the conformal window \cite{athenodorou-1, athenodorou-2}.
This work extends these existing calculations by including the composite fermion and studying whether it becomes a massless free Dirac fermion in the zero quark mass limit.

\section{Background and Previous Work}
The Lagrangian for $SU(2)$ adjoint is
\begin{equation}
\mathcal{L} = \bar\psi (x) (i \slashed{D} - m)\psi(x) - \frac{1}{2} \text{Tr}(G_{\mu\nu}G^{\mu\nu}) 
\end{equation}
where $\slashed{D} = \gamma^\mu(\partial_\mu + i g T^a A^a_\mu(x))$ is written in terms of the adjoint generators $T^a$ and $G_{\mu\nu} = (\partial_\mu A^a_\nu - \partial_\nu A^a_\mu + g f^{abc}A_\mu^b  A_\nu^c) \tau^a$ is written in terms of the fundamental generators $\tau^a$ of $SU(2)$ \cite{athenodorou-1}.

Like QCD, the adjoint representation of $su(2)$ is 3-dimensional, so the invariant tensors are $\delta_{ab}$ and $\epsilon_{abc}$.  Unlike QCD, the adjoint representation is real, so diquark ($\psi \psi$) states as well as $\psi\bar \psi$ states are allowed.  The spectrum therefore consists of the following particles \cite{senthil, athenodorou-1}:
\begin{itemize}[noitemsep,topsep=0pt]
    \item glueballs, functions of the gluonic tensor $G_{\mu\nu}$
    \item diquarks, $\psi_a^T C \Gamma \psi_a$
    \item mesons, $\bar \psi_a \Gamma \psi_a$
    \item hybrid quark-gluon state, $\sigma_{\mu \nu} G_a^{\mu \nu} \psi_a$ 
    \item Dirac composite fermion, $[(\bar \psi_a \psi_b) \gamma_5 \psi_c - (\bar \psi_a \gamma_5 \psi_b) \psi_c] \epsilon_{abc}$
\end{itemize}
where $C=i\gamma_2 \gamma_4$ is the charge conjugation matrix and $\Gamma$ represents some product of Dirac $\gamma$ matrices.

\subsection{Possible Scenarios}
The particle spectrum in the massless limit is constrained by 't Hooft anomaly matching.  In the UV theory, there are 't Hooft anomalies involving the center symmetry and the discrete chiral flavor symmetry residual after symmetry breaking from the $U(1)_A$ anomaly, which must be reproduced in the IR theory.  This requires the presence of massless particles with the correct quantum numbers.  At least three possible scenarios satisfy these anomaly matching constraints \cite{scenarios}:
\begin{itemize}[noitemsep,topsep=0pt]
    \item Conformality: When the quark mass vanishes, the theory becomes scale-free, and all hadrons and glueballs become massless.  For small finite quark masses, hadron masses scale like $m_q^{1/(1+\chi)}$, where $\chi$ is the conformal dimension of the theory (the same for all particles).  One could also have a weaker version of conformality where some particles are gapped out but a nontrivial \emph{conformal sector} becomes massless.
    \item Dirac fermion conjecture: When $m_q \rightarrow 0$, the composite fermion becomes massless and non-interacting, while the remainder of the spectrum could either be gapped or gapless.
    \item Chiral symmetry breaking: Analogously to QCD, chiral symmetry spontaneously breaks when there is no explicit breaking by the quark mass term.  
    As a result, the Goldstone bosons (here the scalar diquark $\psi^T C \gamma_5 \psi$ and its antiparticle, the ``pions'' of this theory) become massless while the rest of the spectrum is gapped out.  Near the massless limit, the scalar diquark has mass given by the Gell-Mann--Oakes--Renner relation, $m_\pi \propto m_q^{1/2}$ \cite{gmor}.
\end{itemize}
Sufficiently precise measurements of the energy spectrum for small $m_q$ and a range of lattice spacings could distinguish between these scenarios.


\subsection{Previous Calculations}
The previous lattice simulations of 1-flavor $SU(2)$ adjoint \cite{athenodorou-1, athenodorou-2} measured the diquarks, mesons, glueballs, and the hybrid but not the composite fermion.  While the hybrid and composite fermion interpolators have nonzero overlap at nonzero quark mass, it is unclear \emph{a priori} how strong this overlap is. Moreover, the overlap between these states vanishes in the chiral limit. Therefore, the measurements of the hybrid particle might not be reliable predictions for the composite fermion.

The bulk of the measurements in \cite{athenodorou-1, athenodorou-2} were taken at a single value of the gauge coupling constant ($\beta=2.05$, defined in Ref.~\cite{athenodorou-1}), with a small number of particles also measured at $\beta=2.20$.  At $\beta=2.05$, all particle masses scaled like $m_q^{1/(1+\chi)}$ with $\chi \approx 0.95$, consistent with the conformality conjecture.  The lightest measured particle was the scalar meson, which the authors identified with the dilaton of the conformal (or near-conformal) theory and as a candidate Higgs boson \cite{athenodorou-1}.

This work extends previous calculations by including the composite fermion in the spectrum. Future work will extend these calculations to finer lattice spacings.

\section{Gauge Field Generation}
To compute the gauge action, the gauge field is stored as $2\times 2$ matrices in $SU(2)$, but to apply the Dirac operator to adjoint fermions, the gauge fields must be mapped into the 3-dimensional adjoint representation.  Similarly, during gauge generation, the fermion forces computed must be converted back into $2\times 2$ matrices.  
The calculations in this work used the USQCD software package Qlua \cite{qlua}, designed to combine the flexible code development of the Lua scripting language and the high performance of the QDP and QUDA libraries \cite{quda}, in order to efficiently generate gauge fields with these non-standard routines included in fermion force calculations.
For this one-flavor theory, the Rational Hybrid Monte Carlo (RHMC) algorithm was used to generate the gauge configurations \cite{rhmc}.
Qlua routines for force computation in RHMC were written and linked to the QUDA multishift inverter to accelerate the calculations.  The coefficients of the rational expansion were generated using a standalone implemention of the Remez algorithm \cite{remez}.

In order to reproduce the results of \cite{athenodorou-1} as a cross-check of the code written here, this work used gauge coupling $\beta=2.05$ and unimproved Wilson fermions with bare quark masses ranging from $-1.475$ to $-1.524$.  Lighter quark masses hit the Aoki phases, so studies using finer lattice spacing or an improved action are needed to explore the chiral limit.  For this preliminary calculation, lattice volumes were $12^3 \times 24$, but in all cases, $m_\pi L > 4.8$.  (Many of the calculations in \cite{athenodorou-1} used larger lattice volumes, but comparing results showed no noticeable finite volume effects.)

\section{Spectroscopy}
This work measured the diquarks, mesons, hybrid, and composite fermion but not the glueballs (which have already been computed \cite{athenodorou-1}).  Signal quality was improved by smearing the source and sink with wavefunction and link smearing
.  Computation of the diquarks is straightforward since the correlator is fully connected; the other particles are discussed below.

\subsection{PCAC Mass}
The bare quark mass is additively renormalized for Wilson fermions, but the partially conserved axial current (PCAC) mass is a measurement of explicit chiral symmetry breaking due to quark masses that vanishes in the chiral limit.  
For large $t$, the PCAC mass is defined as
$$ m_\text{PCAC}(t)=\frac{\langle \partial_t A_4(t) P(0) \rangle}{\langle P(t)P(0) \rangle}$$
with $A_4(t) = \bar\psi(t) \gamma_4 \gamma_5 \psi(t)$, $P(t)=\bar\psi(t) \gamma_5 \psi(t)$ \cite{athenodorou-1}.  In this work, the time derivative is defined by fitting $A_4(t)P(t)$ to a hyperbolic sine and differentiating analytically.

\subsection{Hybrid Fermion}
The hybrid fermion is comprised of a single quark and gluons, which can be combined in a gauge-invariant way in $SU(2)$ adjoint as $\sigma_{\mu \nu} G_a^{\mu \nu} \psi_a$, where $\sigma_{\mu \nu} = \frac{1}{2}[\gamma_\mu, \gamma_\nu]$, and $G_{\mu \nu}$ is a sum of plaquettes with the same structure as a clover term \cite{athenodorou-1}.  Since $\sigma_{\mu \nu} G_a^{\mu \nu}(x) \psi_a(x)$ is gauge invariant, the wall source $\Sigma_{\vec{x}} \sigma_{\mu \nu} G_a^{\mu \nu}(x) \psi_a(x)$ can be constructed without the need for gauge fixing.

\subsection{Composite Fermion and Mesons}
\begin{figure}
    \centering
    \includegraphics[width=\textwidth]{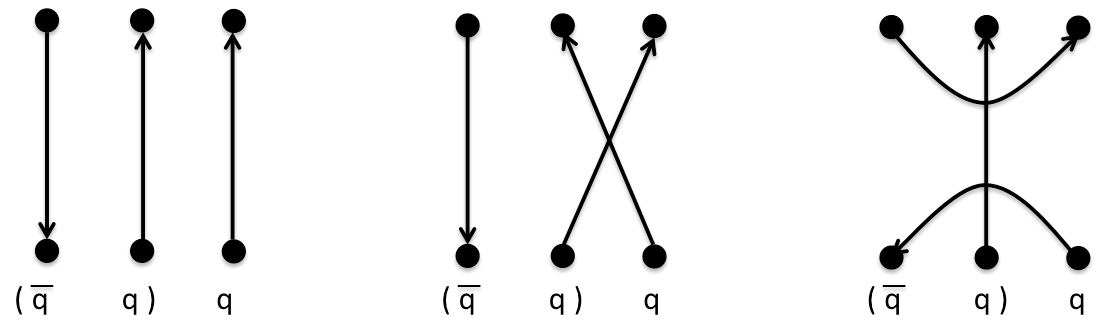}
    \caption{The nonzero contractions necessary to compute the composite fermion.  The quark-antiquark pair in parenthesis show the grouping that is represented in the composite correlator $[(\bar \psi_a \psi_b) \gamma_5 \psi_c - (\bar \psi_a \gamma_5 \psi_b) \psi_c] \epsilon_{abc}$, where the antiquark is paired with one of the quarks.  The contractions not shown vanish by charge conjugation symmetry.}
    \label{contractions}
\end{figure}

The mesons, as in QCD, contain connected and disconnected components.  
The composite fermion contains two connected diagrams and one partially-disconnected diagram (see Figure \ref{contractions}).

The connected piece could be computed relatively cheaply, but the partially disconnected diagrams are expensive.
The disconnected loops in both the composite fermion and the mesons can be computed either by using the Hutchinson trace \cite{hutchinson-1, hutchinson-2} 
or via the sparsening procedure described below.  
Empirically, with unsmeared correlators, the sparsening method gave a better quality signal per inversion for the composite fermion.  A more thorough comparison that includes the effects of smearing will follow in future work. 

\subsection{Sparsening and Contractions}

The composite fermion interpolating operator can be expressed in terms of a contraction of three fermion fields with a weight tensor analogous to the weight tensors used in constructing baryon and multi-baryon interpolating operators in lattice QCD~\cite{Basak:2005ir,Doi:2012xd,blocks},
\begin{equation}
\begin{split}
f_\sigma(x) = \left[(\overline{\psi}_a(x) \psi_b(x))\gamma_5 - (\overline{\psi}_a(x) \gamma_5 \psi_b(x)) \right] P_\sigma \psi_c(x) \epsilon_{abc} = \sum_\alpha w^{(\sigma)\alpha}_{ijk} \overline{\psi}_i(x) \psi_j(x) \psi_k(x),
\end{split}
\end{equation}
where $i,j,k = 1,...,12$ indexes the components of the tensor product of the Dirac spinor and adjoint color algebras, 
$P_\sigma$ projects out the fermion component with positive parity and spin $\sigma = \pm 1/2$, and $\alpha$ indexes the non-zero weights. 
 In terms of the weights and fermion propagators $S_{i^\prime i}(x,y) = \left< \psi_{i^\prime}(x) \overline{\psi}_i(y)\right>$, composite fermion correlation functions are given by 
\begin{equation}
\begin{split}
    &\left< f_\sigma(t,\mathbf{p})\overline{f}_\sigma(0,\mathbf{p}) \right> = \sum_{\alpha, \alpha'} \sum_{\mathbf{x},\mathbf{y}} w^{(\sigma)\alpha^\prime}_{i^\prime j^\prime k^\prime}w^{(\sigma)\alpha}_{i j k} e^{i\mathbf{p}\cdot (\mathbf{x} - \mathbf{y})} \\
    &\hspace{10pt} \times \left\lbrace  \gamma_5 S_{i^\prime j}(\mathbf{x},t;\mathbf{y},0)^\dagger \gamma_5 \left[ S_{k^\prime i}(\mathbf{x},t;\mathbf{y},0) S_{j^\prime k}(\mathbf{x},t;\mathbf{y},0) - S_{j^\prime k}(\mathbf{x},t;\mathbf{y},0) S_{i^\prime i}(\mathbf{x},t;\mathbf{y},0) \right] \right. \\
    &\hspace{20pt} - \left. S_{j k}(\mathbf{x},0;\mathbf{y},0)\left[  S_{k^\prime i}(\mathbf{x},t;\mathbf{y},0) S_{j^\prime i^\prime}(\mathbf{x},t;\mathbf{y},t) - S_{j^\prime i}(\mathbf{x},t;\mathbf{y},0) S_{i^\prime k^\prime}(\mathbf{x},t;\mathbf{y},t) \right] \right.\\
      &\hspace{20pt} + \left. S_{j i}(\mathbf{x},0;\mathbf{y},0)\left[  S_{k^\prime k}(\mathbf{x},t;\mathbf{y},0) S_{j^\prime i^\prime}(\mathbf{x},t;\mathbf{y},t) - S_{j^\prime k}(\mathbf{x},t;\mathbf{y},0) S_{i^\prime k^\prime}(\mathbf{x},t;\mathbf{y},t) \right] \right\rbrace
    \label{eq:weights}
    \end{split}
\end{equation}

Due to the terms involving equal-time propagators associated with the partially disconnected diagrams in Fig.~\ref{contractions}, evaluating this expression requires propagators from each source point to itself and thus requires an approximation to the all-to-all propagator.  This can be made tractable by making use of the observation in Ref.~\cite{coarsening} that hadron correlators can be computed on a sparse grid of sink points.
If the same sparse grid of points is used for the source and sink positions, then all-to-all fermion propagators on the sparse lattice can be computed tractably. 
This can be used to compute any hadron correlator, whether it be fully connected, partially disconnected, or fully disconnected.
In particular the composite fermion correlator is computed simply by restricting the lattice site sums in Eq.~\eqref{eq:weights} to the sparse lattice. 
This leads to additional excited-state contamination from higher momentum modes not exactly projected out of the correlation function, but this can be controlled through Euclidean time evolution analogously to other excited-state effects \cite{coarsening}.

\section{Results and Future Work}
\begin{figure}
    \centering
    \begin{subfigure}[t]{0.48\textwidth}
        \centering
        \includegraphics[width=\textwidth]{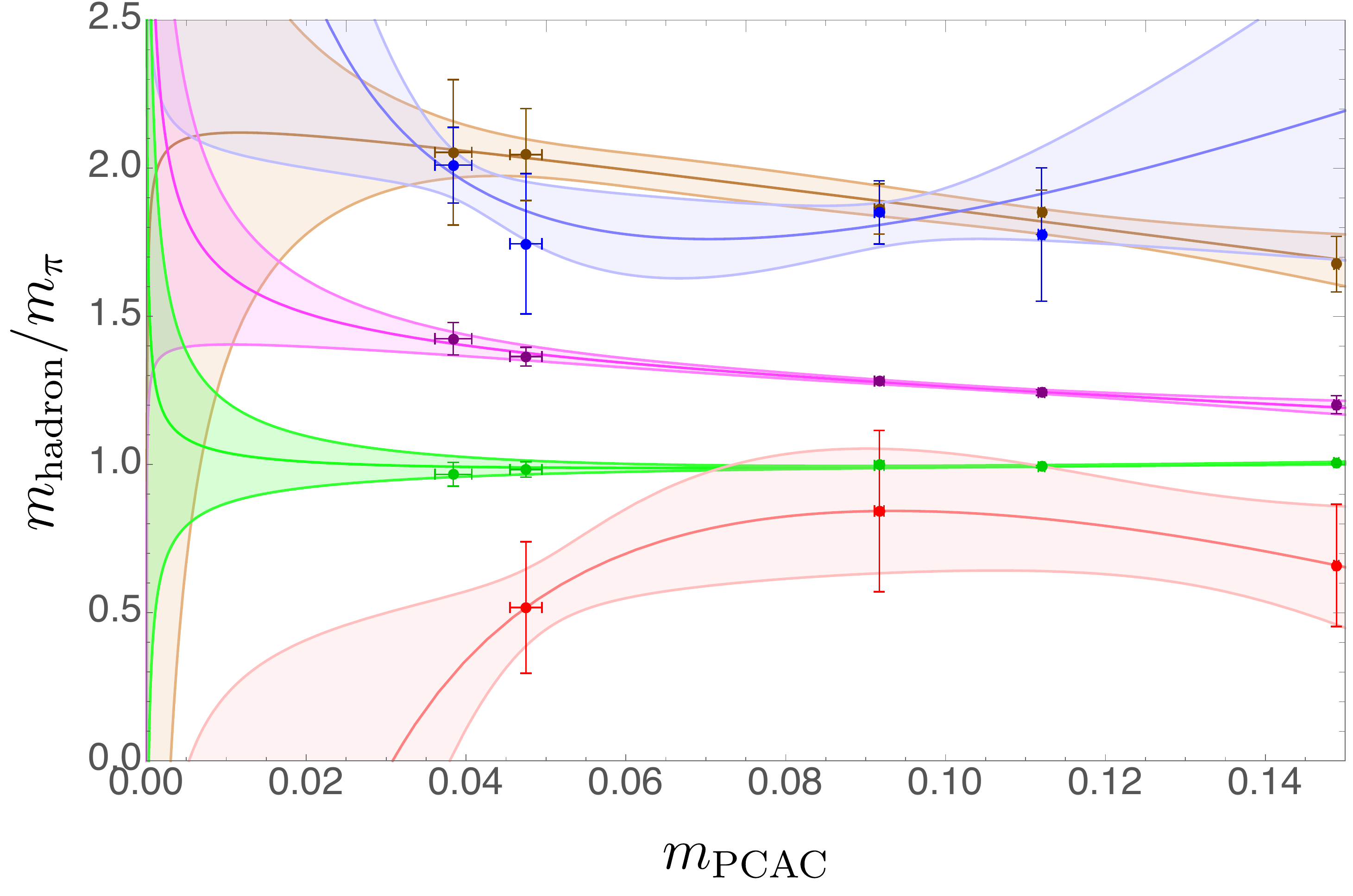}
        \caption{A plot of the masses of (from heaviest to lightest) the \textcolor{hybrid}{hybrid} and \textcolor{composite}{composite} fermions and the \textcolor{vector}{vector}, \textcolor{pseudoscalar}{pseudoscalar}, and \textcolor{scalar}{scalar} mesons, all normalized by dividing by the pion (scalar diquark) mass.  Scalar meson masses are from \cite{athenodorou-1}; all other masses were measured here.  The fits have the functional form $a m_\text{PCAC} + b/\sqrt{m_\text{PCAC}} + c$, which is general enough to fit any of the three hypotheses investigated.}
        \label{ratios}
    \end{subfigure}
    ~
    \begin{subfigure}[t]{0.48\textwidth}
        \centering
        \includegraphics[width=\textwidth]{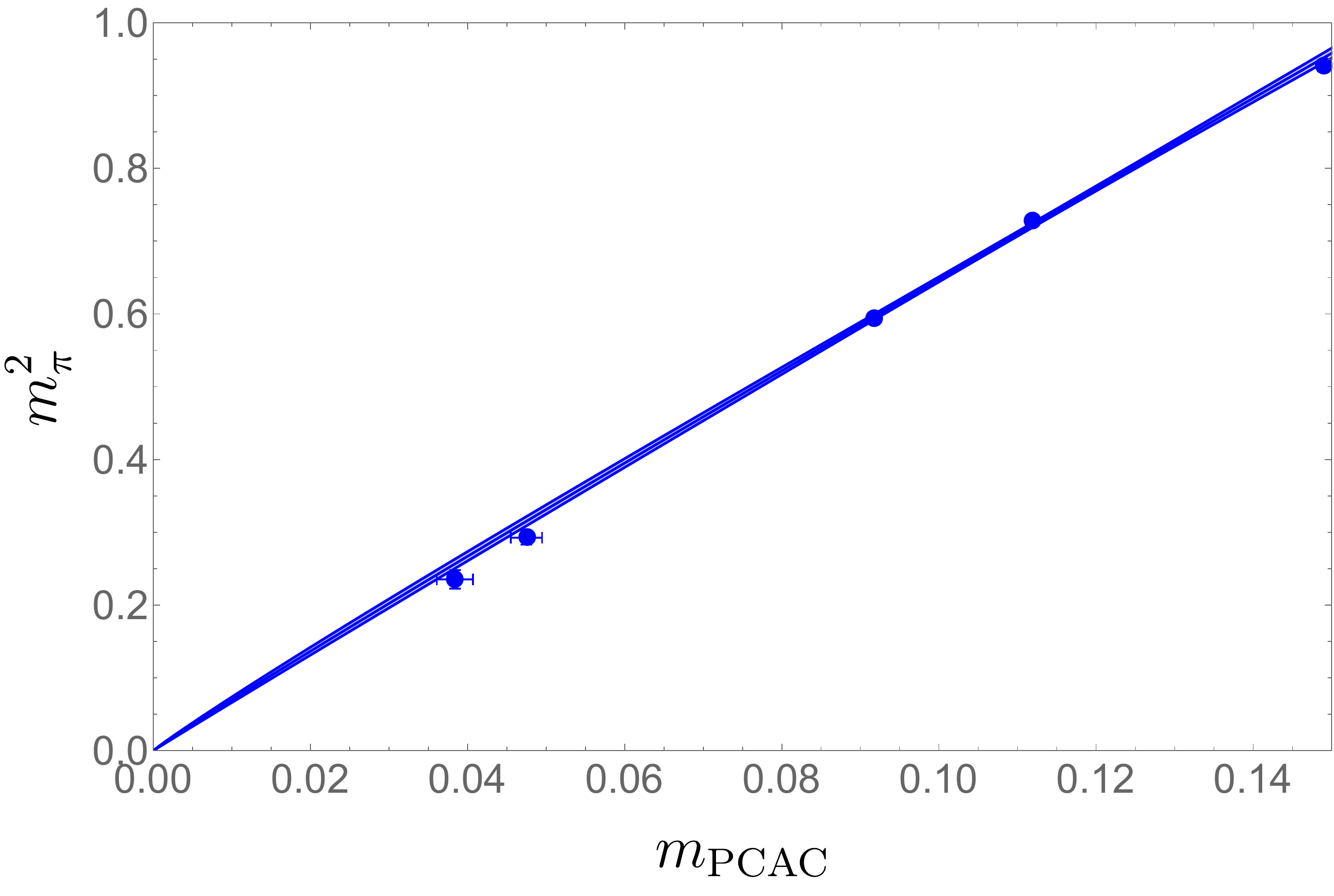}
        \caption{A plot of $m_\pi^2$ versus $m_\text{PCAC}$ with a power-law fit.  
        The conformal hypothesis predicts power-law scaling of the pion mass with the quark mass; the $\chi$SB hypothesis also predicts power law scaling but with a specific power ($m_\pi^2 \propto m_q$, which would appear linear on the plot).  The power-law fit gives an exponent that does not exclude either $\chi$SB or the conformal scenario suggested by Ref.~\cite{athenodorou-1}.}
        \label{chipt}
    \end{subfigure}
\end{figure}

The mass measurements agree with \cite{athenodorou-1} for all particles measured in both calculations (namely, the scalar and pseudoscalar diquarks, the vector and pseudoscalar mesons, and the hybrid fermion).
The masses of several particles in the spectrum, all normalized by dividing by the pion mass, are shown in Figure \ref{ratios}.  The conformal hypothesis predicts that these ratios are independent of the PCAC mass, which is consistent with data measured here for all particles except the vector meson\footnote{It is difficult to say from these ensembles alone whether deviation from conformality of the vector meson is a physical effect or an artifact of the relatively coarse lattice spacing.  Even if physical, however, one cannot exclude the possibility of a conformal sector containing the remainder of the particle spectrum.}.

The chiral symmetry breaking hypothesis makes the additional prediction that $m_\pi^2 \propto m_\text{PCAC}$, and the data are consistent with this relation as well (Figure \ref{chipt}).  Chiral symmetry breaking also implies that the pion should be lighter than all other states as $m_q \rightarrow 0$, while the scalar meson appears to be lighter than the pion over the mass range studied.  However, these data only cover a limited range of quark masses, so either the pion or the composite fermion could become lighter than the remainder of the spectrum as quark mass decreases further.  Thus, it is not possible to exclude any of the three scenarios with this data set, necessitating the need for further study.

Differentiating between these scenarios will require increasing measurement precision and also collecting data at lighter quark masses, closer to the chiral limit.  Furthermore, since these simulations were all performed at a fixed value of $\beta$, it will be necessary to repeat this process at finer lattice spacings to understand discretization effects and to take the continuum limit.

\section{Acknowledgements}
This work is supported in part by the U.S. Department of Energy, Office of Science, Office of Nuclear Physics, under grant Contract Number DE-SC0011090, and by the SciDAC4 award DE-SC0018121. ZB and MLW are also supported by MIT Pappalardo fellowships.  The authors thank Will Detmold, Phiala Shanahan, Andrew Pochinsky, and Senthil Todadri for useful discussions.

\end{document}